\documentclass[12pt,a4in]{article}
\usepackage{graphicx}
\usepackage{geometry}
\usepackage[francais,english]{babel}
\begin{document}
\setcounter{page}{0}
\thispagestyle{empty}

\setcounter{page}{1}

\renewcommand{\thesection}{\Roman{section}}
\begin{center}
\large{\bf Application of supersymmetric quantum mechanics to study bound state properties of  exotic hypernuclei }\\
\vspace{0.5cm}
\normalsize\it
Md. Abdul Khan\\
\footnotesize
Department of Physics, Aliah University, \\
AA-II, Plot No. IIA/27, Newtown, Kolkata-700156, West Bengal, India.\\
Email: drakhan.rsm.phys@gmail.com; drakhan.phys@aliah.ac.in\\
\end{center}
\begin{center}

\begin{abstract}
Bound state properties of few single and double-$\Lambda$ hypernuclei is critically examined in the framework of core-$\Lambda$  and core+$\Lambda+\Lambda$ few-body  model applying hyperspherical harmonics expansion method (HHEM). The $\Lambda\Lambda$ potential is chosen
phenomenologically while the core-$\Lambda$ potential is 
obtained by folding a phenomenological $\Lambda N$ interaction into the
density distribution of the core. The depth of the effective $\Lambda N$
potential is adjusted to reproduce the experimental data for the
core-$\Lambda$ subsystem. The three-body Schr\"odinger equation is solved by
hyperspherical adiabatic approximation (HAA) to get the 
ground state energy and wave function.  The ground state wavefunction is used to construct the supersymmetric partner potential following prescription of supersymmetric quantum
mechanics (SSQM) algebra. The newly constructed supersymmetric partner potential is used to solve the three-body Schr\"odinger equation to get the energy and wavefunction for the first excited state of the original potential. The method is repeated to predict energy and wavefunction  of the next higher excited states. The possible number of bound states is found to increase with the increase in mass of the core of the hypernuclei. The Root Mean Squared (RMS) matter radius and some other relevant geometrical observables are also predicted.
\end{abstract}
\end{center}
\vspace{1cm}
PACS \hspace{45pt}21.80.+a; 21.60.Jz; 21.30.Fe\\
\vspace{0.5cm}
\newpage
\rm

\newpage
\section{\bf Introduction}
Since the discovery of the light exotic hypernuclei in the early sixties 
[1-2], peoples are giving much attention to study the structure of such
multistrange hypernuclei [3-4]. Some of the experimentally observed one- 
and two-$\Lambda$ hypernuclei are $_{\Lambda}^{5}$He,
$_{\Lambda\Lambda}^{6}$He, $_{\Lambda}^{9}$Be, 
$_{\Lambda\Lambda}^{10}$Be,  $_{\Lambda}^{13}$C and 
$_{\Lambda\Lambda}^{13}$B [5-11]. Information about the $\Lambda\Lambda$
and $\Lambda N$ forces can be extracted from the bound state properties of
such nuclei. These informations on $\Lambda\Lambda$ and $\Lambda N$
interaction together with the well known $NN$ interaction may give better
understanding of the  strong quark-quark (qq) interaction,
since the hyperons also have qqq structure as that of nucleons. As for
example proton has three quarks namely up,up,down ($p\rightarrow uud$),
neutron has quarks up,down,down ($n\rightarrow udd$), $\Lambda^0$ has
quarks up,down,strange ($\Lambda^0\rightarrow uds$) etc. In the earlier
stage of their discovery, emulsion experiments provided some information
on the binding energies of $\Lambda$-particle in the light exotic
hypernuclei and their decay rates (life times$\sim 10^{-10} sec$) [2].
From the available binding energy data physicists gathered some 
qualitative informations about the $\Lambda$-nucleon ($\Lambda N$)
interaction and single particle potential strength for the
$\Lambda$-particle in hypernuclei [12]. The hyperon-nucleon scattering
experiments have also been performed but these are not so profound and are
still in the primary stages and do not give detailed phase shifts to
construct the potential reliably. Some $\Lambda N$ and $\Sigma N$ total
cross-sections and very few angular distribution at low energies have been
measured [13-18], but these are not sufficient to allow the phase shift
analysis. Under the circumstances, the bound state properties of single-
and double- $\Lambda$ hypernuclei can only give useful indirect
information about $\Lambda N$ and $\Lambda\Lambda$ interactions.
For example, one can take phenomenological forms of  
$\Lambda N$ and $\Lambda\Lambda$ interactions and see if they reproduce the
experimentally observed observables of the hypernuclei. Alternatively one
can adjust the parameters of the empirical potential to reproduce the
bound state properties and thus predict the effective $\Lambda N$ and
$\Lambda\Lambda$ interactions. 
In some earlier works [19-22] physicists have used variational 
and approximate few body methods for the hypernuclei treating them 
a few body system.
In the present work, we investigate the structure and properties of the
ground and excited states of double-$\Lambda$ hypernuclei with
mass number $A=6,8,34,42,92$ and ground state of their binary subsystems
consisting the core and one outer core $\Lambda$ hyperon.
A simple two-body model has been adopted here for the single-$\Lambda$
hypernucleus ($A=A_c+1$) consisting of a $A_c=4,6,30,40,90$ core and one
valence $\Lambda$ particle. And for the double-$\Lambda$ hypernuclei
($A=A_c+2$), we adopt a three-body model consisting core
of mass $A_{c}$ (=4,6,30,40,90) and two valence $\Lambda$ particles.
For $\Lambda\Lambda$ potential a phenomenological form has been
chosen. The core-$\Lambda$ potential has been
obtained by folding an effective $\Lambda N$ potential by the
density of the core nucleus. The strength of the 
effective $\Lambda N$ potential for a given core is adjusted to reproduce
the experimentally known BE of the single $\Lambda$ hypernuclei 
$_{\Lambda}^{5}$He, $_{\Lambda}^{7}$He [6], $_{\Lambda}^{7}$Be [23],
$_{\Lambda}^{33}$S [24], $_{\Lambda}^{41}$Ca [24] 
and $_{\Lambda}^{91}$Zr [25]. 
The same folded core-$\Lambda$ potential is then used for the
double-$\Lambda$ hypernucleus for the same core.\\
We used hyperspherical harmonics expansion (HHE) method to solve the
three-body system. This method is a powerful tool for the ${ab~ initio}$
solution of the few body Schr\"{o}dinger equation for a given set of
binary interaction potentials among the constituent particles. This method
has been used for bound states in atomic [26-33], nuclear [34-40] and particle
physics [41-43]. Attempts have been made to use it in scattering problems
as well [44]. In this method, the wave function is expanded in a complete
set of hyperspherical harmonics (HH), which are, for a three-body system,
the six-dimensional analogue of ordinary spherical harmonics. The
Schr\"{o}dinger equation reduces to a set of coupled differential 
equations which can be solved numerically. The HHE method is essentially
exact and involves no approximation other than an eventual truncation of
the expansion basis. Any desired precision in the binding energy can, in
principle, be achieved by gradually expanding the expansion basis and
checking the rate of convergence. However the number of coupled
differential equations and hence, the complexity in the numerical solution
increases rapidly as the expansion basis is increased by introducing
larger hyper angular momentum quantum numbers. Computer limitations set an
ultimate limit to the precision attainable.\\
We have calculated the two-$\Lambda$ separation energy 
($B_{\Lambda\Lambda}$), $\Lambda\Lambda$ bond energy ($\Delta
B_{\Lambda\Lambda}$) and some size parameters for all the above double
$\Lambda$ hypernuclei. Here $B_{\Lambda\Lambda}$ and $\Lambda\Lambda$ bond
energy $\Delta{B_{\Lambda\Lambda}}$ are defined as  
\begin{equation}
B_{\Lambda\Lambda}(_{\Lambda\Lambda}^{A}Z) = [M(^{A-2}Z)+ 2
M_{\Lambda} - M(_{\Lambda\Lambda}^{A}Z)] c^{2} 
\end{equation}
and
\begin{equation}
\Delta{B_{\Lambda\Lambda}}=B_{\Lambda\Lambda}(_{\Lambda\Lambda}^{A}Z)
-2B_{\Lambda}(_{\Lambda}^{A-1}Z)
\end{equation}
 As the ground state BE is quite large, one may expect the excited states 
to be observable experimentally, although they have not been reported so 
far. However, the HHE method is known for
its rather slow convergence, which is especially manifested in the excited
states. The excited state wave function is more spread out than the ground
state and therefore needs larger $K_{max}$ for the same degree of
convergence. We circumvent this difficulty by invoking supersymmetric
quantum mechanics (SSQM) [45]. In SSQM, we construct the supersymmetric
partner potential ($V_{2}$) of the original potential ($V_{1}$) [1].
According to SSQM, the spectra of $V_{1}$ and $V_{2}$ are identical except
for the absence of the ground state of $V_{1}$ in the spectrum of $V_{2}$
[45]. Hence the ground state of $V_{2}$ corresponds to the first
excited state of $V_{1}$. Thus we can solve for the ground state of
$V_{2}$, whose convergence is fast (being a ground state). Its BE is the
BE of the first excited state of the original potential. The wave function
of the first excited state of $V_{1}$ can be obtained by applying an
appropriate operator on the ground state wave function of $V_{2}$.\\
 The paper is organized as  
follows: In section II, we review the HHE method for a three-body system
consisting of non identical particles with a brief review of SSQM.
Results of calculation and discussion are presented in section III.
Finally in section IV we draw our conclusions. 

\section{\bf Use of hyperspher spherical harmonics expansion method and super
symmetric quantum mechanics} We treat each of the six
double-$\Lambda$ hypernuclei $_{\Lambda\Lambda}^6$He,
$_{\Lambda\Lambda}^8$Be, $_{\Lambda\Lambda}^8$He,
$_{\Lambda\Lambda}^{34}$S, 
$_{\Lambda\Lambda}^{42}$Ca, $_{\Lambda\Lambda}^{92}$Zr as a three
body system, where each of the core $^{4}$He, $^{6}$Be, $^{6}$He,
$^{32}$S, $^{40}$Ca, $^{90}$Zr is labeled as particle 1 and the two
valence $\Lambda$ hyperons as particles 2 and 3 respectively (see Fig. 1).

\begin{figure}
\centering
\fbox{\includegraphics[width=0.85\linewidth, height=0.65\linewidth]{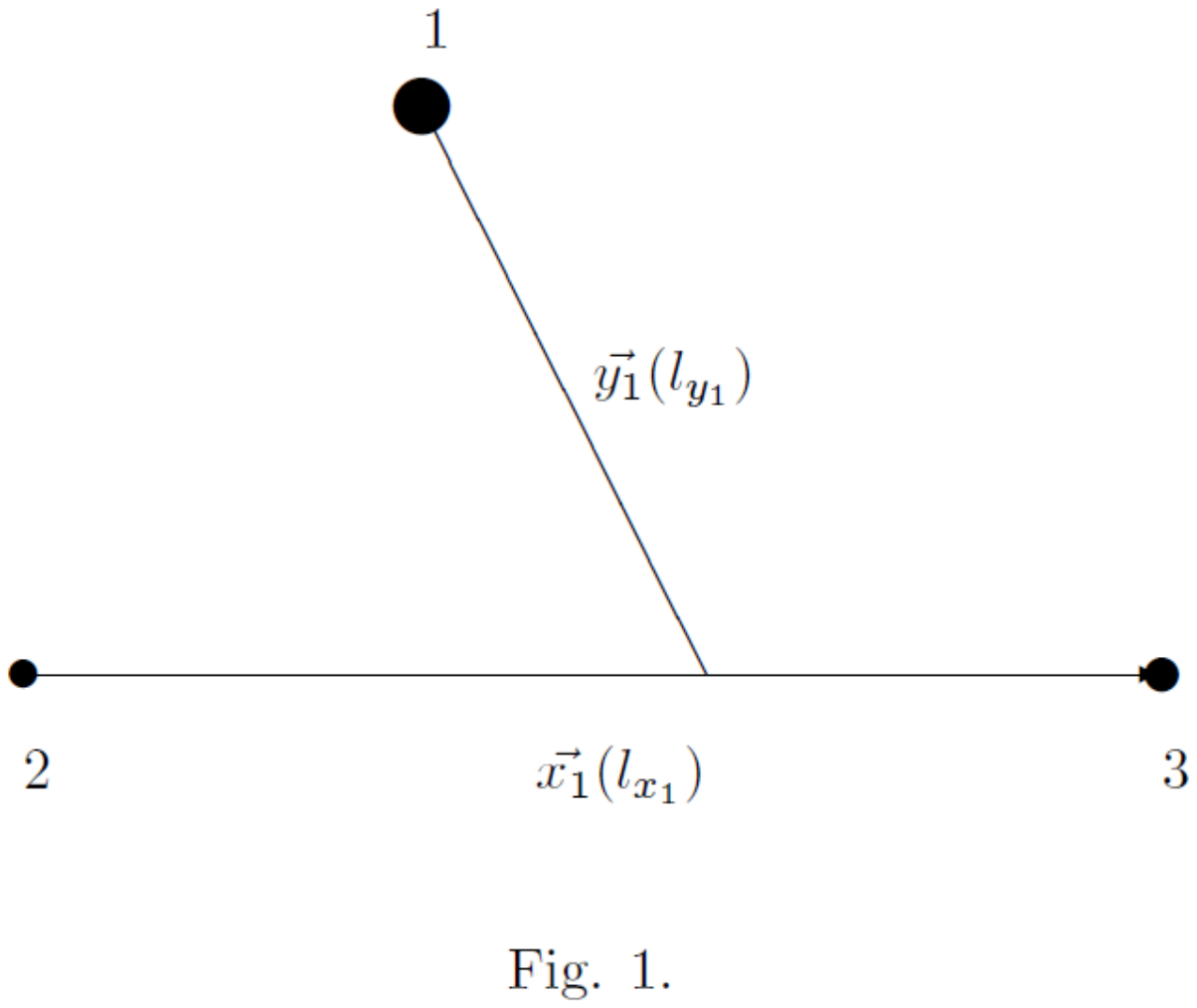}}
\caption{Choice of Jacobi coordinates for the partition 1.}
\label{fig:boxed_graphic}
\end{figure}

 For a given partition $i$ (in which $i$ is the spectator), a set of
Jacobi co-ordinates is defined as:
\begin{equation}
 \left.  \begin{array}{ccl}
   \vec{x_{i}} & = & \left[\frac{m_{j}m_{k}M}{m_{i}(m_{j}+m_{k})^{2}}
\right]^{\frac{1}{4}} (\vec{r_{j}} - \vec{r_{k}}) \\ 
   \vec{y_{i}} & = & \left[\frac{m_{i}(m_{j}+m_{k})^{2}}{m_{j}m_{k}M} 
\right]^{\frac{1}{4}} \left(\vec{r_{i}} - \frac{m_{j}
\vec{r_{j}} + m_{k} \vec{r_{k}}}{ m_{j} + m_{k}} \right)  \\
	 \vec{R} & = & \frac{1}{M} \left(m_{i} \vec{r_{i}} + m_{j} 
\vec{r_{j}} + m_{k} \vec{r_{k}} \right).
	   \end{array}  \right\} 
\end{equation}
for (i,j,k)=(1,2,3 cyclic). Here $m_{i}$, $\vec{r_{i}}$ are the mass and
position of the $i^{th}$ particle and $M = m_{i} + m_{j} + m_{k}$.
$\vec{R}$ is the coordinate of the centre of mass of the system. 
Since the interactions among the constituent particles depend only on
their relative separations, the centre of mass motion separates out
automatically. Thus the relative motion of the three-body system is
described by the Schr\"{o}dinger equation 
\begin{equation}
\left[ - \frac{\hbar^{2}}{2\mu} (\nabla_{x_{i}}^{2} + \nabla_{y_{i}}^{2})+ 
V_{jk} (\vec{x_{i}})
 +V_{ki} (\vec{x_{i}}, \vec{y_{i}} ) + V_{ij} (\vec{x_{i}}, \vec{y_{i}} )
-E \:  \right] \Psi ( \vec{x_{i}}, \vec{y_{i}} ) \:= \: 0
\end{equation}
where ${\mu\: =\: \left[ \frac{m_{i} m_{j} m_{k}}{M}
\right]^{\frac{1}{2}}}$ is an effective mass parameter and $V_{ij}$ is the
interaction potential between particles $i$ and $j$. In terms of
the hyperspherical variables [40]
\begin{equation}
 \left. \begin{array}{ccl}
 x_{i} & = & \rho \cos \Phi_{i} \\
 y_{i} & = & \rho \sin \Phi_{i}
	  \end{array} \right\} 
\end{equation}
the Schr\"{o}dinger equation becomes 
\begin{equation}
\left[ - \frac{\hbar^{2}}{2\mu} \{ \frac{1}{\rho^{5}}
\frac{\partial}{\partial\rho}(\rho^{5}  \frac{\partial}{\partial\rho}) -
\frac{\hat{{\cal K}}^{2}(\Omega_{i})}{\rho^{2}} \} + V ( \rho, \Omega_{i}
) - E \right] 
\Psi (\rho, \Omega_{i} ) \:=\: 0,
\end{equation}
where the hyper-radius $\rho$ is invariant under three-dimensional
rotations and permutations of the particle indices. 
The five hyper angles constituted by polar angles
($\theta_{x_{i}},\phi_{x_{i}}$ of $\vec{x_i}$), polar 
angles ($\theta_{y_{i}},\phi_{y_{i}}$) of $\vec{y_i}$ and 
the angle $\Phi_i$,  are collectively denoted by
$\Omega_i\rightarrow\left\{\Phi_i,\theta_{x_i},\theta_{y_i},\phi_{x_i},
\phi_{y_i})\right\}$. Thus ($\rho, \Omega_i$) constitutes the six
hyperspherical variables of which the five hyper angles $\Omega_i$ depend
on the choice of partition $i$. In the above equation 
$V(\rho, \Omega_{i})$ = $ V_{jk}(\vec{x_{i}}) +
V_{ki}(\vec{x_{i}}, \vec{y_{i}}) + V_{ij}(\vec{x_{i}}, \vec{y_{i}})$ is
the total interaction potential expressed in terms of the hyperspherical 
variables in the partition $i$ and $\hat{{\cal K}}^{2}(\Omega_{i})$ is the
square of hyper angular momentum operator given by [40]
\begin{equation}
\hat{{\cal K}}^{2}(\Omega_{i}) \:=\: - \: \frac{\partial^{2}}{{\partial
\Phi_{i}}^{2}} - 
 4 \cot 2\Phi_{i}~ \frac{\partial}{\partial \Phi_{i}} + \frac{1}{\cos^{2}
\Phi_{i}} \hat{l}^{2} (\hat{x_{i}}) + \frac{1}{\sin^{2}
\Phi_{i}} \hat{l}^{2} (\hat{y_{i}}),
\end{equation}
where $\hat{l}^{2} (\hat{x_{i}})$ and $\hat{l}^{2} (\hat{y_{i}})$ are the 
squares of ordinary orbital angular momentum operators associated with
$\vec{x_{i}}$ and $\vec{y_{i}}$ motions respectively. The operator
$\hat{{\cal K}}^{2}$ satisfies an eigenvalue equation [40] 
\begin{equation}
\hat{{\cal K}}^{2}(\Omega_{i}) {\cal Y}_{K \alpha_{i}}(\Omega_{i}) \:=\: K
(K+4)  {\cal Y}_{K \alpha_{i}}(\Omega_{i}).
\end{equation}
Here K ($= 2n_{i} + l_{x_{i}} + l_{y_{i}}, n_i$ being a non-negative
integer) represents the hyper angular momentum quantum number and 
 $\alpha_{i}\equiv{\{ l_{x_{i}}, l_{y_{i}}, L, M \} }$ is a short
hand notation, in which L and M denote the total orbital angular momentum 
and its projection, respectively. The normalized eigen functions
${\cal Y}_{K\alpha_{i}}(\Omega_{i})$ are called the hyperspherical
harmonics (HH) and these depend on the choice of partition. A detailed
analytic expression for ${\cal Y}_{K\alpha_{i}}(\Omega_{i})$ can be found
in [40].
In the HHE method, $\Psi(\rho, \Omega_{i})$ is expanded in the complete
set of HH corresponding to the partition $i$: 
\begin{equation}
\Psi(\rho, \Omega_{i}) = \sum_{K\alpha_{i}}\frac{U_{K\alpha_{i}}   
  (\rho)}{\rho^{5/2}} {\cal Y}_{K\alpha_{i}}(\Omega_{i}).
\end{equation}
Substitution of Eq. (9) in Eq. (6) and the use of orthonormality of HH leads
to a set of coupled differential equations (CDE) in $\rho$ 
\begin{equation}
\begin{array}{cl}
& \left[ -\frac{\hbar^{2}}{2\mu} \left( \frac{d^{2}}{d\rho^{2}}
-\frac{{\cal L}_{K} ({\cal L}_{K}+1)}{\rho^{2}}\right) - E \right] 
U_{K \alpha_{i}}(\rho)  \\
+ & \sum_{K^{\prime} \alpha_{i}^{~\prime}} < K \alpha_{i}
\mid V(\rho, \Omega_{i}) \mid K^{\prime} \alpha_{i}^{~\prime}
> U_{K^{\prime} \alpha_{i}^{~\prime}}(\rho) \: = \: 0,
\end{array}
\end{equation}
where ${{\cal L}_{K} = K + 3/2 }$ and
\begin{equation}
<K \alpha_{i} | V (\rho, \Omega_{i}) | K^{\prime} \alpha_{i}^{~\prime} > =
{\displaystyle{\int}}_{\Omega_{i}} 
{\cal Y}_{K\alpha_{i}}^{*}(\Omega_{i}) V(\rho, \Omega_{i}) {\cal
Y}_{K^{\prime} 
 \alpha_{i}^{~\prime}}(\Omega_{i}) d\Omega_{i}.
\end{equation}
The number of coupled differential equations to be solved, Eq. (10) is in 
principle, an infinite one, which arises out of an infinite number of
basis states Eq. (9). For practical computation, the hyperspherical
harmonics expansion basis in Eq. (9) is truncated to a finite set by
retaining all values of K up to a maximum $K_{max}$. For a given K, all
allowed values of the set of quantum numbers $\alpha_i$ are included. The 
basis states is further reduced by imposing the constraints arising out of
symmetry requirements and associated conserved quantum numbers. Thus the
infinite set of CDE is now reduced to a finite set.
This finite set of CDE is solved numerically by hyperspherical adiabatic
approximation [31,46] which is based on
the idea of adiabatic separation of hyper-angular motions from the
hyper-radial one. In this procedure the potential matrix 
$<K\alpha_{i}|V|K^{\prime}\alpha_{i}^{\prime}>$ together with the diagonal
hyper-centrifugal term is diagonalized for a fixed value of $\rho$. The
lowest eigenvalue $\omega_{0}(\rho)$ and the corresponding eigenvector 
$\chi_{K\alpha_{i}, 0}(\rho)$ are obtained as parametric function of
$\rho$. Then the set of CDE, Eq. (10) is approximately decoupled into a
single ordinary differential equation
\begin{equation}
\left[ - \frac{\hbar^{2}}{2\mu}  \frac{d^{2}}{d\rho^{2}}
+ \omega_{0}(\rho)-E \right] \psi(\rho) =0
\end{equation}
Numerical solution of Eq. (12), subject to appropriate boundary conditions
gives the ground state energy E and partial waves of the three-body
system. The partial waves are given by 
\begin{equation}U_{K\alpha_{i}}(\rho)\simeq 
\psi_{0}(\rho)\chi_{K\alpha_{i}, 0}(\rho)
\end{equation}
 We now present a brief review of properties of supersymmetric
quantum mechanics (SSQM) [45] used here to calculate the observables for
the excited states.
Let us consider a one dimensional Schr\"{o}dinger equation of the form 
\begin{equation}
\left[ - \frac{\hbar^{2}}{2\mu}  \frac{d^{2}}{d\rho^{2}}
+ V(\rho)-E \right] \psi(\rho) =0
\end{equation}
For the single-$\Lambda$ hypernuclei, $V(\rho)$ is just the $\Lambda$-core
potential plus the centrifugal term and $\rho$ is the core-$\Lambda$
separation. For double-$\Lambda$ hypernuclei $V(\rho)$ is the lowest eigen
potential $\omega_{0}(\rho)$ and $\rho$ is the hyper-radius. Suppose 
the ground state energy is $E_{0}$. We shift the energy scale by $E_{0}$,
such that the ground state is at zero energy and the corresponding
potential is renamed $V_{1}(\rho)$ (=${V(\rho)-E_{0}}$). Then for the
ground state 
 \begin{equation}
\left[ - \frac{\hbar^{2}}{2\mu}  \frac{d^{2}}{d\rho^{2}}
+ V_{1}(\rho)\right] \psi_{0}(\rho) =0
\end{equation}
From this we get 
\begin{equation}
V_{1}(\rho)=\frac{\hbar^{2}}{2\mu}\frac{\psi_{0}^{\prime\prime}(\rho)}
{\psi_{0}(\rho)} 
\end{equation}
Let us now define a super potential, $W(\rho)$ as   
\begin{equation}
W(\rho)=-\frac{\hbar}{\sqrt{2\mu}}\frac{\psi_{0}^{\prime}(\rho)}
{\psi_{0}(\rho)} 
\end{equation}
and two operators
\begin{equation}
\left. \begin{array}{lcl}
A&=&\frac{\hbar}{\sqrt{2\mu}}\frac{d}{d\rho}+W(\rho)\\
A^{\dagger}&=&-\frac{\hbar}{\sqrt{2\mu}}\frac{d}{d\rho}+W(\rho) 
       \end{array} \right\}
\end{equation}
Then one can easily check that 
\begin{equation}
V_{1}(\rho)=W^{2}(\rho)-\frac{\hbar}{\sqrt{2\mu}} W^{\prime}(\rho)
\end{equation}
and
\begin{equation}\begin{array}{lllll}
H_{1}&\equiv - \frac{\hbar^{2}}{2\mu}  \frac{d^{2}}{d\rho^{2}}
+ V_{1}(\rho)&=&A^{\dagger}A
                \end{array}        
\end{equation}
Here $H_{1}$ is the original Hamiltonian in the shifted energy scale.
We next define a partner Hamiltonian
\begin{equation}\begin{array}{lllll}
H_{2}&=&AA^{\dagger}&=&\frac{\hbar^{2}}{2\mu}  \frac{d^{2}}{d\rho^{2}}
+ V_{2}(\rho),   
                \end{array}        
\end{equation}
in which the partner potential is given by 
\begin{equation}
V_{2}(\rho)=W^{2}(\rho)+\frac{\hbar}{\sqrt{2\mu}} W^{\prime}(\rho).
\end{equation}
Let us denote the $n^{\underline{th}}$ eigenvalue and corresponding eigen
function of $H_{2}$ by $E_{n}^{(2)}$ and $\psi_{n}^{(2)}(\rho)$
respectively, while the $n^{\underline{th}}$ eigenvalue and corresponding
eigenfunction of $H_{1}$ are denoted by $E_{n}$ and $\psi_{n}(\rho)$
respectively. Then it is easily verified that 
\begin{equation}\begin{array}{lllllll}
H_{2}(A\psi_{n})&=&AA^{\dagger}A\psi_{n}&=&AH_{1}\psi_{n}&=
&E_{n}(A\psi_{n}),
                \end{array}        
\end{equation} 
which means that $A\psi_{n}$ is an eigenfunction of $H_{2}$ corresponding
to energy $E_{n}$. Similarly
 \begin{equation}\begin{array}{lllllll}
H_{1}(A^{\dagger}\psi_{n}^{(2)})&=&A^{\dagger}AA^{\dagger}\psi_{n}^{(2)}&
=&A^{\dagger}H_{2}\psi_{n}^{(2)}&=&E_{n}^{2}(A^{\dagger}\psi_{n}^{(2)})
                \end{array}        
\end{equation} 
Thus $A^{\dagger}\psi_{n}^{(2)}$ is an eigenfunction of $H_{1}$
corresponding to energy $E_{n}^{(2)}$. Now from Eq. (17) and (18) 
\begin{equation} A\psi_{0}=0. \end{equation}
Thus $A\psi_{0}$ is a trivial solution of Eq. (23). The ground state of
$H_{2}$ will thus be $A\psi_{1}$ corresponding to energy $E_{1}$ (note
that ${E_{0}< E_{1}\leq E_{2}\leq E_{3}\leq ......}$). Hence
$E_{0}^{(2)}=E_{1}$, and from Eq. (24), $\psi_{1}\infty 
A^{\dagger}\psi_{0}^{(2)}$.Thus if
we solve for the ground state of $V_{2}(\rho)$, we get the energy $E_{1}$
(in shifted energy scale), which is the energy of the first excited state
of $V_{1}(\rho)$. It is easily seen that the corresponding normalized
eigenfunction is 
\begin{equation}
\psi_{1}(\rho)=\frac{1}{\sqrt{E_{1}}}A^{\dagger}\psi_{0}^{(2)}.
\end{equation}
Hence we numerically solve Eq. (14) for the ground state and calculate the
super-potential $W(\rho)$ and the partner potential $V_{2}(\rho)$ according
to Eq. (17) and (22) respectively. Then we solve the same Schr\"{o}dinger
equation with $V_{2}(\rho)$ to obtain $E_{0}^{(2)}$ and $\psi_{0}^{(2)}$.
From $\psi_{0}^{(2)}$, we get $\psi_{1}(\rho)$ using eq(26). Finally
the energy of the first excited state of $V_{1}(\rho)$ is obtained by
back shifting $E_{0}^{(2)}$. 
\section{\bf Results and Discussions}
In the present calculation we have taken the core to be
structureless. Since the core ($^{4}$He, $^{6}$Be, $^6$He, $^{32}$S,
$^{40}$Ca, $^{90}$Zr) contains only nucleons and no $\Lambda$-particles,
there is no symmetry requirements under exchange of the valence $\Lambda$
particles with the core nucleons. The only symmetry requirements are (i)
anti-symmetrization of the core wave function under exchange of the
nucleons and (ii) anti-symmetrization of the three body wave function under
exchange of the two $\Lambda$ particles. The former is implicitly taken
care of by the choice of the core as a building block. For
double-$\Lambda$ hypernuclei, the latter is
correctly incorporated by restricting the $l_{x_{1}}$ values, explained  
in the following. Thus, within the three-body model, the
symmetry requirements are correctly satisfied without any approximation.
For the single-$\Lambda$ hypernuclei, there are no symmetry requirements.
The ground state of all experimentally known double-$\Lambda$
hypernuclei have a total angular momentum quantum number $J=0$ and
even (or positive) parity. We 
assume this to be true for all double-$\Lambda$ hypernuclei with core
having $N=even, Z=even$. The possible total spin (S) of the three-body
system $(core+\Lambda+\Lambda)$ can take two values 0 or 1 since the spin
of the core in all the above cases has a value 0. Thus the total orbital 
angular momentum L can be either 0 or 1 corresponding to S=0 or 1
respectively. Hence the set of quantum numbers (LS)J for the ground state 
of all even-even core double-$\Lambda$ hypernuclei is (00)0 and (11)0 
which  corresponds to $^1S_0$ and $^3P_0$ states. Since the ground state 
of these nuclei have definite parity, odd and even parity states will not
mix. In our case, the ground state of all the six double-$\Lambda$ 
hypernuclei is a pure $^1S_0$ state. 
Since the core is spin less, the spin singlet state (S=0) corresponds to
zero total spin of the valence $\Lambda$- particles (i.e. $S_{23}$=0).
Hence the spin part of the wave function is antisymmetric under the
exchange of the spins of the two $\Lambda$-particles. Thus the spatial
part must be symmetric under the exchange of the two $\Lambda$-hyperons.
The symmetry of the spatial part is determined by the hyper spherical
harmonics, since the hyper radius $\rho$ and hence the hyper radial
partial waves ($U_{K\alpha}(\rho)$) are invariant under permutation of the
particles. Under the pair exchange operator $P_{23}$ which interchanges
particles 2 and 3, ${\vec{x_{1}}\rightarrow \vec{-x_{1}}}$ and
$\vec{y_{1}}$ remains unchanged (see Eq. (3)). Consequently $P_{23}$ acts
like the parity operator for (23) pair only. Choosing the two valence
$\Lambda$-hyperons to be in spin singlet state (spin antisymmetric), the
space wave function must be symmetric under $P_{23}$. This then requires
$l_{x_{1}}$ to be even. For the spin singlet state total orbital angular
momentum, L=0, hence we must have ${l_{x_{1}} = l_{y_{1}}}$ = even
integer. Since ${K=2n_{1}+l_{x_{1}}+l_{y_{1}}}$, where $n_{1}$ is a
non-negative integer, $K$ must be even and 
\begin{equation}
\left. \begin{array}{rclcl}
 l_{x_{1}}&=&l_{y_{1}}&=& 0,2,4,..,K/2~~~~ if~ K/2 ~is~ even  \\
&&&&0,2,4,..,(K/2-1)~~ if~ K/2~ is~ odd
       \end{array} \right\}.
\end{equation}
Again for the triplet state (S=1), the two valence $\Lambda$-hyperons
will be in spin triplet state ($S_{23}=1$, spin symmetric). Hence the
space wave function must be antisymmetric under $P_{23}$. This then
requires $l_{x_{1}}$  to be odd. For the spin triplet state, the total
orbital angular momentum, L=1, hence $l_{y_{1}}$ may take values
$l_{x_{1}}$ and ${l_{x_{1}}\pm 1}$ (and only 1 if $l_{x_{1}}=0$) 
but the parity conservation allows
$l_{y_{1}}$=$l_{x_{1}}$ only (except $l_{x_{1}}=0$). Again since
${K=2n_{1}+l_{x_{1}}+l_{y_{1}}}$, where $n_{1}$ is a non-negative integer,
K must be even ($K\neq=0$) and 
\begin{equation}
\left. \begin{array}{rclcl}
 l_{x_{1}}&=&l_{y_{1}}&=& 1,3,5,..,K/2~~~~ if~ K/2 ~is~ odd  \\
&&&&1,3,5,..,(K/2-1)~ if~ K/2~ is~ even
       \end{array} \right\}.
\end{equation}
For a practical calculation, the HH expansion basis (eq(9)) is truncated 
to a maximum value ($K_{max}$) of K. For each allowed K $\le K_{max}$
with K=even integers, all allowed values of $l_{x_{1}}$ are included.
The even values of $l_{x_{1}}$ 
correspond to L=0, S=0 and odd values of $l_{x_{1}}$ correspond to L=1,
S=1. This truncates eq(10) to a set of N coupled differential equations,
where 
\begin{equation}
\left. \begin{array}{rcl}
N & =& \left(\frac{K_{max}}{2}+1\right)~
\left(\frac{K_{max}}{4}+1\right) ~~ if~ K_{max}/2~ is~ even \\
&&\left(\frac{K_{max}+2}{4}\right)~ \left(\frac{K_{max}}{2}+2 \right)~~
if~ 
K_{max}/2 ~is~ odd 
       \end{array}  \right\},
\end{equation}
which will be solved by the hyperspherical adiabatic approximation (HAA) [46].\\\\

\noindent Since realistic $\Lambda\Lambda$ potential is not available at this
stage. We used phenomenological three term Gaussian $\Lambda\Lambda$ 
potential with parameters adjusted to reproduce the experimental data 
for known double-$\Lambda$ hypernuclei.
The $\Lambda\Lambda$ potential used here is given by  
\begin{equation}
V_{\Lambda\Lambda}(r)=\sum_{i=1}^{3}V_{i}
\exp(-\frac{r^{2}}{\beta_{i}^{2}}).
\end{equation}
The parameters of this potential are presented in Table I. A short range
repulsive term in the $\Lambda\Lambda$ potential is included to simulate
Pauli principle between the valence $\Lambda$ particles. The
core-$\Lambda$ potential is obtained by folding phenomenological
$\Lambda$-nucleon potential into the density
distribution of the core nucleus. The chosen density distribution function
of the core is given by
\begin{equation}
\rho(r)=\frac{\rho_{0}}{1+\exp(\frac{r-r_c}{a})}
\end{equation}
 with $r_c=r_{0} A_{c}^{1/3}$ fm, 
$a=0.65 fm$, $r_{0}=1.7fm$ (where $r_c$ is termed the half density radius 
and a, the skin thickness). The the density constant $\rho_{0}$ is
determined by the condition  
\begin{equation}
\int \rho(r)d^{3}r=Ac,
\end{equation}
where $A_{c}$ is the mass of the core in units of nucleon mass. The value
of $a$ and $r_{0}$ are chosen following suggestions in the literature [10].
The phenomenological $\Lambda N$ potential is given by 
\begin{equation}
    V_{\Lambda N}(r) = V_{0} \exp(-r^{2}/\eta^{2})  
\end{equation} 
with ${V_{0}}$ and $\eta$ adjusted to reproduce the experimental binding 
energy ($B_{\Lambda}$) of single-$\Lambda$ hypernuclei (see Table II) in 
the core-$\Lambda$ subsystem. Then the core-$\Lambda$ potential is
given by 
\begin{equation}
V_{c\Lambda}(r)=\int \rho (r_{1}) V_{\Lambda N}(|\vec{r_{1}}-\vec{r}|)
d^{3}r_{1}. 
\end{equation}
A two dimensional plot of the $\Lambda\Lambda$, effective $\Lambda N$ and 
core-$\Lambda$ folded potential are shown in Fig. 2. The lowest eigen
values ($\omega_0$) of the three body effective potential is plotted 
against the hyper-radial distance $\rho$ as shown in Fig. 3.\\\\

\begin{figure}
\centering
\fbox{\includegraphics[width=0.85\linewidth, height=0.65\linewidth]{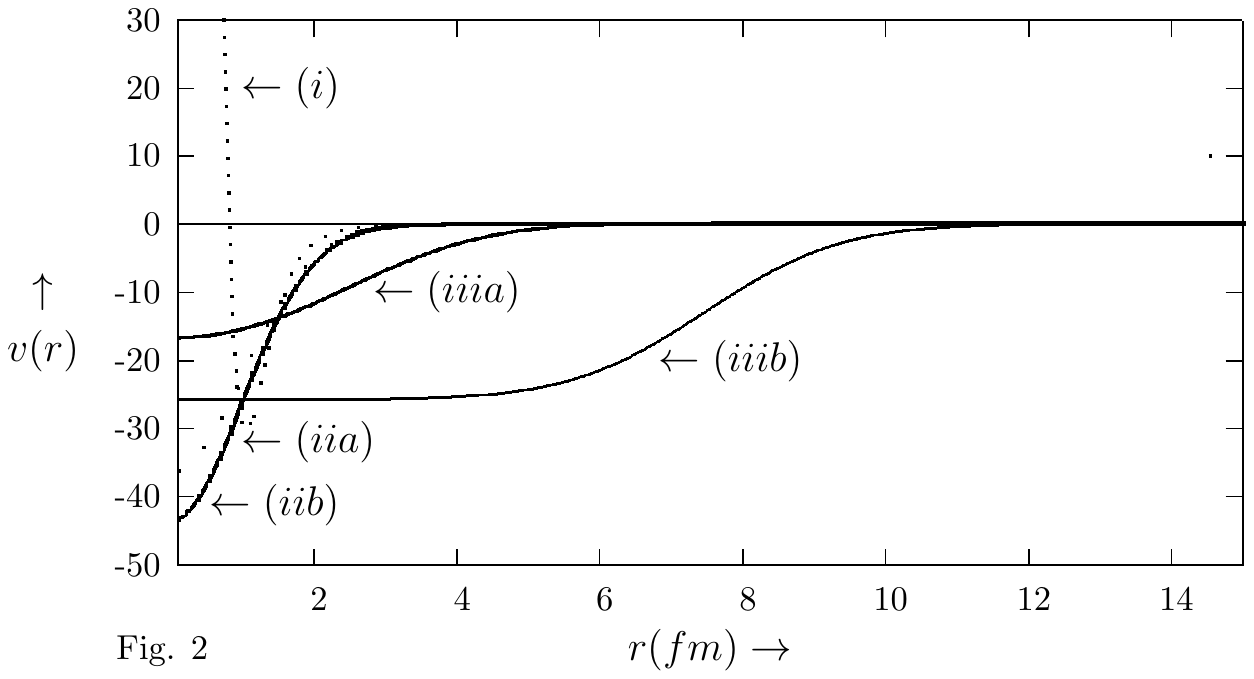}}
\caption{ Plot of: (i) $\Lambda\Lambda$ potential; 
(iia) effective $\Lambda N$ potential for $_{\Lambda\Lambda}^{6}$He;
(iib) effective $\Lambda N$ potential for 
$_{\Lambda\Lambda}^{92}$Zr; (iiia) core-$\Lambda$ folded potential for
$_{\Lambda\Lambda}^{6}$He; 
(iiib) core-$\Lambda$ folded potential for $_{\Lambda\Lambda}^{92}$Zr; 
(v is in MeV and r is the relative separation)}
\label{fig:boxed_graphic}
\end{figure}

\begin{figure}
\centering
\fbox{\includegraphics[width=0.85\linewidth, height=0.65\linewidth]{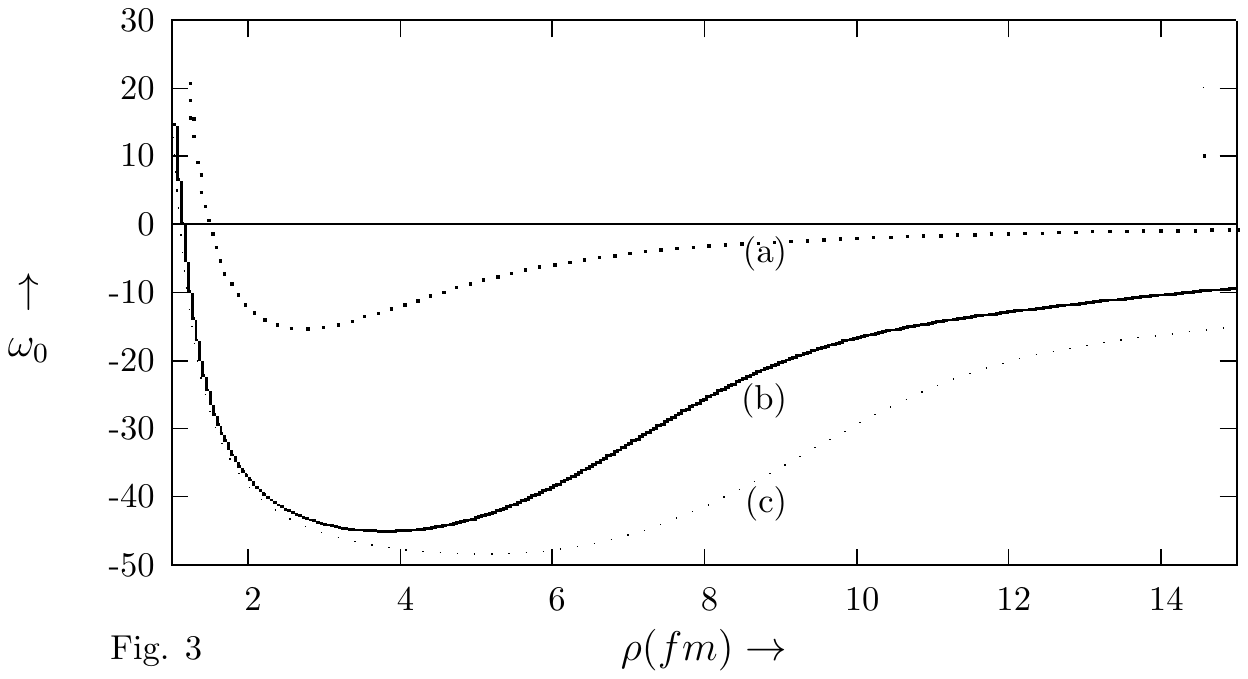}}
\caption{Plot of lowest eigen potential $\omega_0(MeV)$ against hyperradial distance $\rho$ for double-$\Lambda$ hypernuclei 
(a) $_{\Lambda\Lambda}^{6}$He, (b) $_{\Lambda\Lambda}^{42}$Ca and  (c) $_{\Lambda\Lambda}^{92}$Zr}
\label{fig:boxed_graphic}
\end{figure}

The strength of $\Lambda N$ potential is expected to be weakened with
the increase in mass of the core due to the screening or shielding effect
by neighboring nucleons within the core when the interacting nucleon is
embedded in the core. The $\pi$-mesic decay of $\Lambda$ hyperon (${\Lambda
\rightarrow N+\pi}$) is predominant in the free space but tends to be
suppressed in hypernucleus by the Pauli-exclusion principle and instead
nonmesic weak process (${\Lambda + N \rightarrow N +N}$) becomes dominant
with increasing mass number [7,47-51]. Thus we actually get an effective
$\Lambda N$ interaction by the folding process. The parameters of this
effective $\Lambda N$ potential are listed in Table II. The variation
of the depth of the effective $\Lambda N$ interaction against the core 
mass ($A_c$) is displayed in Fig. 4.\\\\

\begin{figure}
\centering
\fbox{\includegraphics[width=0.65\linewidth, height=0.65\linewidth]{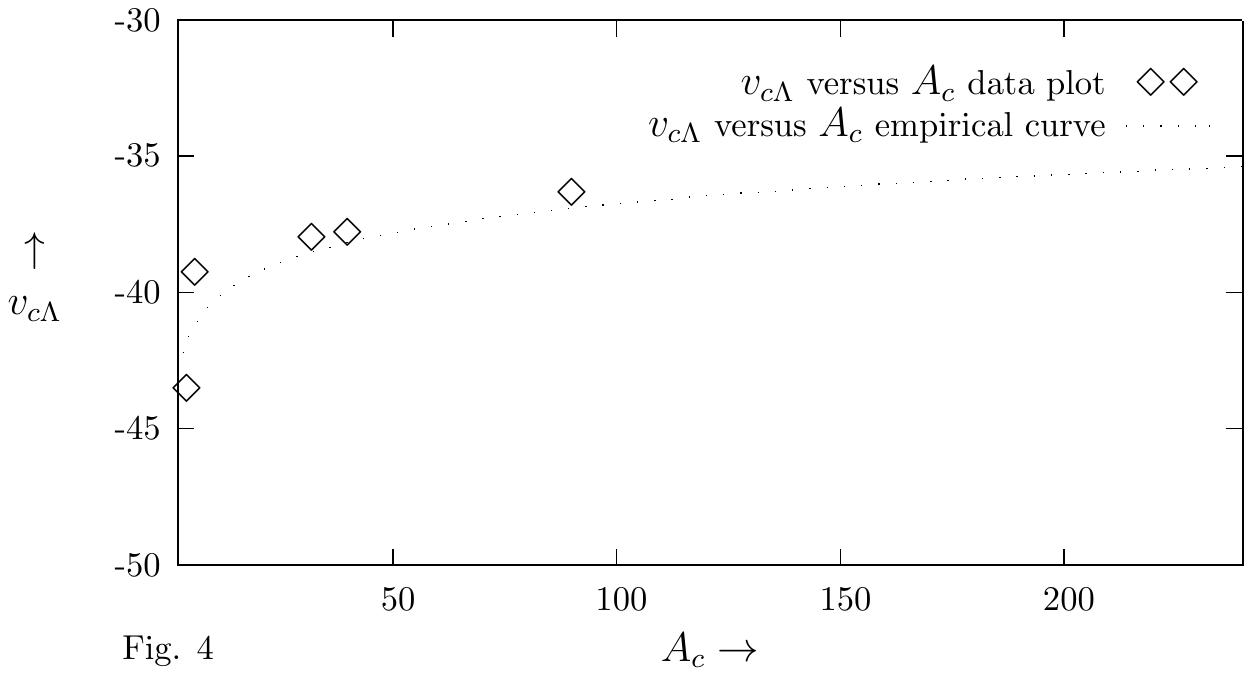}}
\caption{ Plot of core-$\Lambda$ potential depth $v_{c\Lambda}$(MeV) against mass of core $(A_c)$}
\label{fig:boxed_graphic}
\end{figure}
 
The truncated set of CDE for a given $K_{max}$ takes the form 
\begin{equation}
\begin{array}{cl}
& \left[ -\frac{\hbar^{2}}{2\mu} (\frac{d^{2}}{d\rho^{2}}
-\frac{{\cal L}_{K} ({\cal L}_{K}+1)}{\rho^{2}}) - E \right] 
U_{K l_{x_{1}}LS}(\rho)  \\
+ & \sum_{K^{\prime}=0, 2, .... }^{K_{max}}
\sum_{l_{x_{1}}^{\prime}(allowed)}
< K l_{x_{1}} \mid V(\rho, \Omega_{1}) \mid K^{\prime}
l_{x_{1}}^{~\prime}> U_{K^{\prime} 
l_{x_{1}}^{~\prime} L S}(\rho) \: = \: 0 
\end{array}
\end{equation}
(allowed ${l_{x_{1}}^{\prime}}$=0, 2, ... only for S=0, L=0). Note that
the subscripts ${l_{y_{1}}}$ (=$l_{x_{1}}$) or ${l_{y_{1}}^{\prime}}$ 
(=${l_{x_{1}}^{\prime}}$) have been suppressed for brevity. 
Eq. (12) is solved to get the ground state energy $E_{(0)}$ and wave
functions $U_{K\alpha_{1}}^{(0)}$, by the hyperspherical adiabatic
expansion (HAA).  
For the single-$\Lambda$ hypernuclei, we solve Eq. (14) with $V(\rho)$
replaced by $V_{c\Lambda}(\rho)$ plus centrifugal potential. 
The results are presented in the last two columns of Table II. 
For double-$\Lambda$ hypernuclei, we use the same $V_{c\Lambda}$ and 
solve Eq. (12) for the ground state. Then following
the prescription of sec. II, we obtain the energy and wavefunction of
the higher excited states. The ground and excited state wave functions 
for $_{\Lambda\Lambda}^6$He, $_{\Lambda\Lambda}^{42}$Ca, 
and $_{\Lambda\Lambda}^{92}$zr are respectively shown in 
Fig. 5, 6 and 7. The calculated values of 
two-$\Lambda$ separation energy 
($B_{\Lambda\Lambda}$) at $K_{max}=20$ for the ground state (ie. $0^+$), 
different excited states (ie. $0_1^+, 0_2^+, 0_3^+$, etc) and 
$\Lambda\Lambda$ bond energy ($\Delta B_{\Lambda\Lambda}$) for 
the ground state of $_{\Lambda\Lambda}^{6}$He, 
$_{\Lambda\Lambda}^{8}$Be, $_{\Lambda\Lambda}^{8}$He, 
$_{\Lambda\Lambda}^{34}$S, $_{\Lambda\Lambda}^{42}$Ca,
$_{\Lambda\Lambda}^{92}$zr are presented 
in Table III. The variation of the single-$\Lambda$ separation 
energy $B_{\Lambda}$ with the mass of the core $A_c$ is shown 
in Fig. 8. On the same graph, the variation of the two-$\Lambda$ 
separation energy $B_{\Lambda}$ with the mass $A$ is also plotted. 
The calculated two-$\Lambda$ separation energy 
($B_{\Lambda\Lambda}$) and $\Lambda\Lambda$ bond energy ($\Delta
B_{\Lambda\Lambda}$) of $_{\Lambda\Lambda}^{6}$He are in agreement with 
the experimental value 7.25$\pm$0.19 $MeV$ [7] and 1.01 $MeV$ [7]
respectively. The predicted two-$\Lambda$ separation energies for the 
ground and excited states of all the six double$\Lambda$ hypernuclei 
considered here are depicted in the energy level diagram (see Fig. 9).  

\begin{figure}
\centering
\fbox{\includegraphics[width=0.75\linewidth, height=0.6\linewidth]{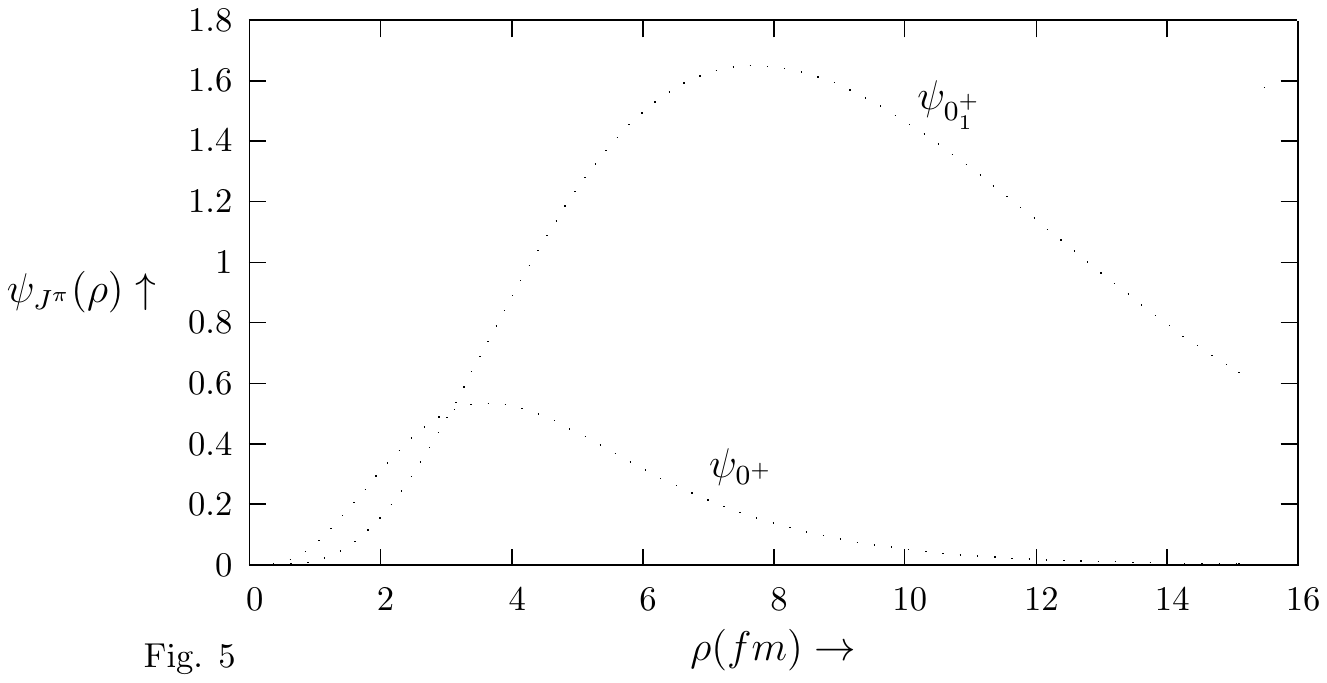}}
\caption{ Plot of ground  and excited state wavefunctions for $_{\Lambda\Lambda}^{6}$He}
\label{fig:boxed_graphic}
\end{figure}

\begin{figure}
\centering
\fbox{\includegraphics[width=0.75\linewidth, height=0.6\linewidth]{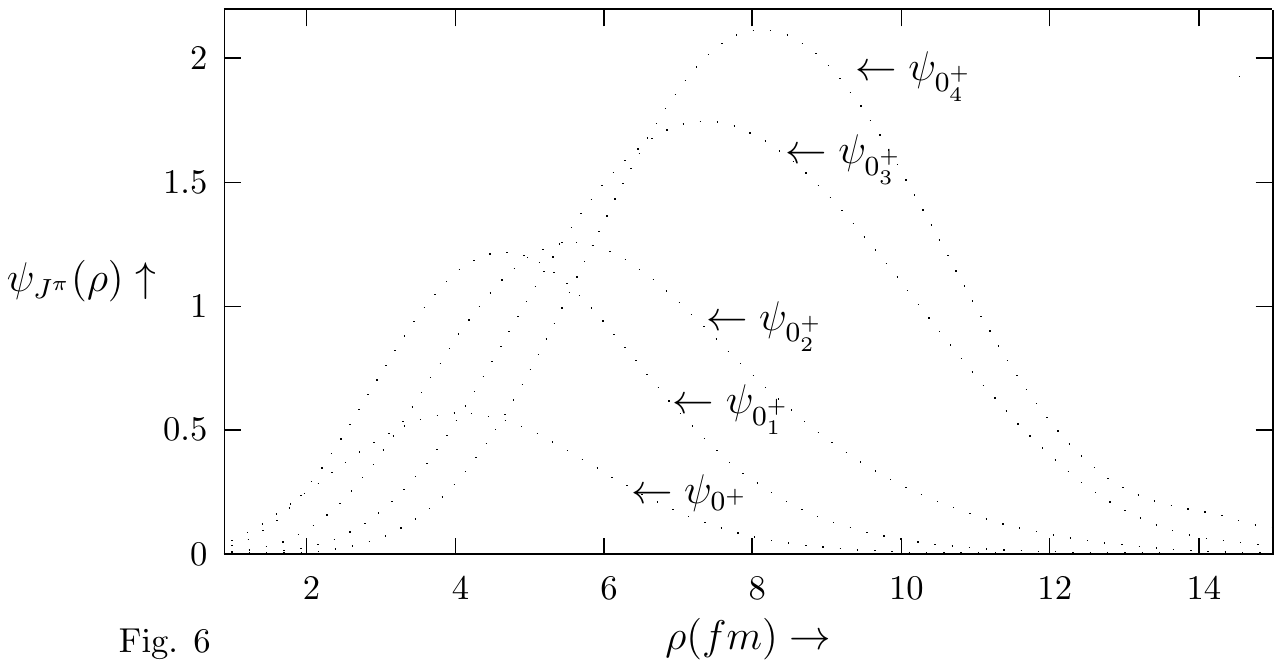}}
\caption{Plot of ground  and excited state wavefunctions for $_{\Lambda\Lambda}^{42}$Ca}
\label{fig:boxed_graphic}
\end{figure}

\begin{figure}
\centering
\fbox{\includegraphics[width=0.75\linewidth, height=0.6\linewidth]{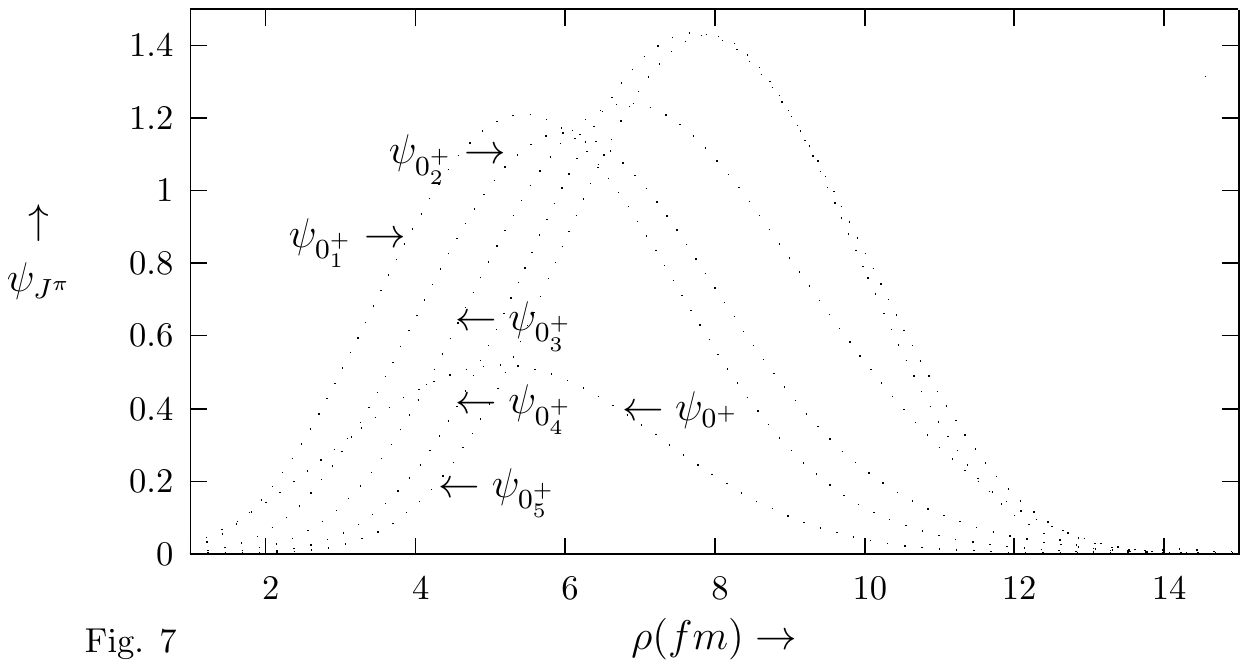}}
\caption{Plot of ground  and excited state wavefunctions for $_{\Lambda\Lambda}^{92}$Zr}
\label{fig:boxed_graphic}
\end{figure}

\begin{figure}
\centering
\fbox{\includegraphics[width=0.75\linewidth, height=0.6\linewidth]{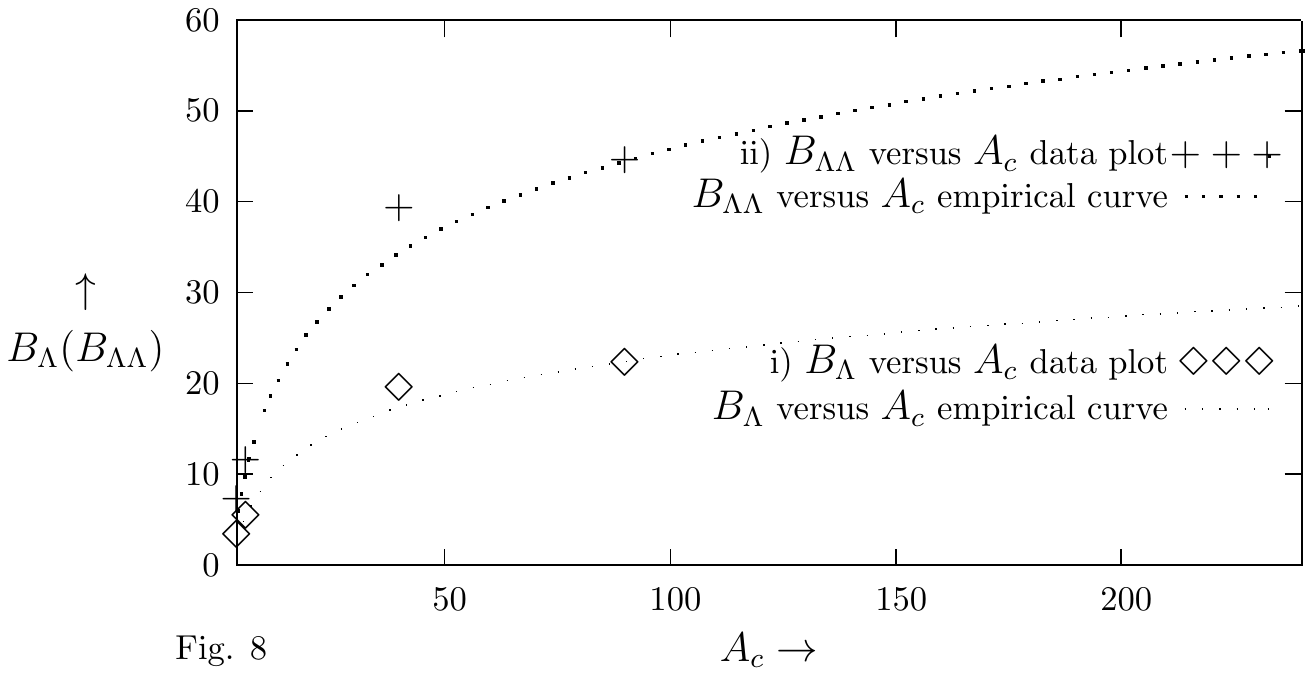}}
\caption{Plot of ground state- (i) Single-$\Lambda$ separation energy $B_{\Lambda}$(MeV) and (ii) Two-$\Lambda$ separation 
energy $B_{\Lambda\Lambda}$ (MeV), against mass of the core $A_c$}
\label{fig:boxed_graphic}
\end{figure}

\begin{figure}
\centering
\fbox{\includegraphics[width=0.75\linewidth, height=0.6\linewidth]{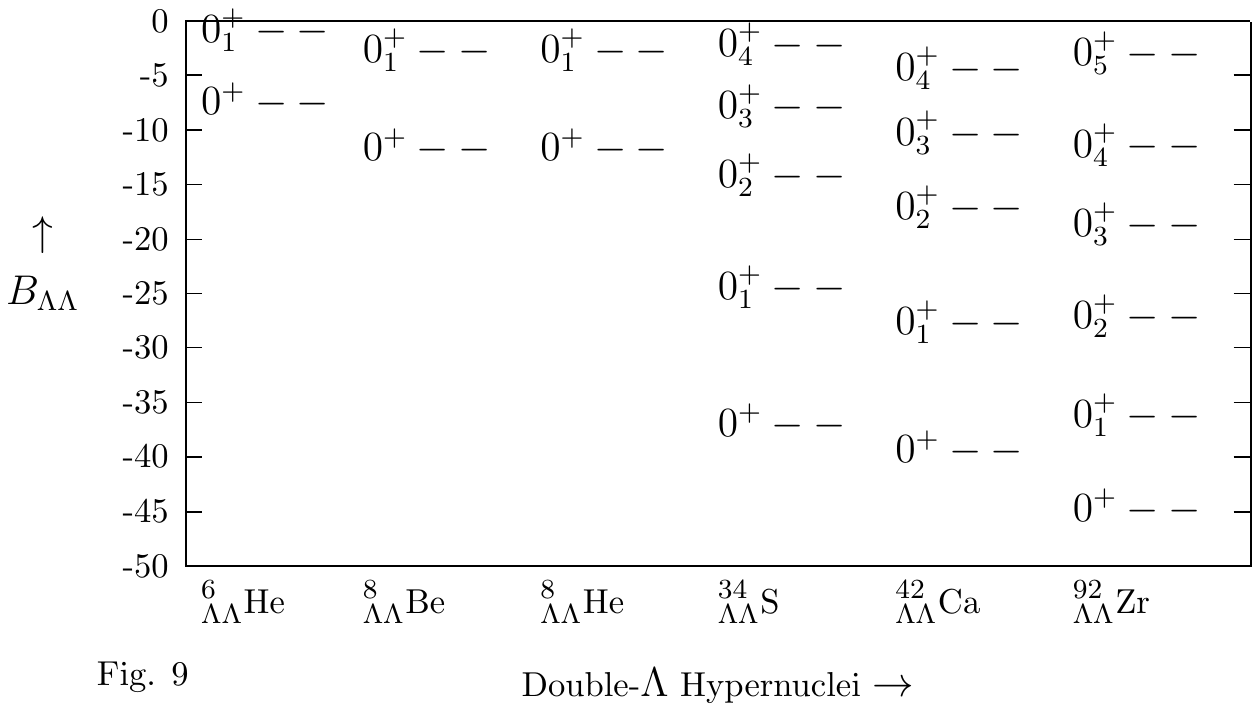}}
\caption{Energy Level diagram of  Double-$\Lambda$ Hypernuclei [$B_{\Lambda\Lambda}$ is in MeV]}
\label{fig:boxed_graphic}
\end{figure}

After obtaining the three body wave function by the HHE approach, 
root mean square (rms) matter radius ($R_M$) of the three-body 
system has been calculated following the equation
\begin{equation}
R_{M}=\left[ \frac{A_{c}R_{c}^2 + m_{\Lambda}<
r_{13}^2+r_{12}^2>}{A_{c}+2m_{\Lambda}} \right]^{1/2}, 
\end{equation}
where$A_{c}$, $m_{\Lambda}$ are the masses of the core and the $\Lambda$
hyperon (in units of nucleon mass) and $R_{c}$ is the matter radius of the
core determined by the relation $R_{c}=r_{0}A_{c}^{1/3}$ with $r_{0}=1.70
fm$. The rms core-$\Lambda$ separation is defined as 
\begin{equation}
R_{c\Lambda}=\left[ \frac{< r_{13}^2 + r_{12}^2 >}{2} \right]^{1/2}.
\end{equation}
The expectation value of the observable ${< r_{13}^2+r_{12}^2 >}$ is
given by the expression 
\begin{equation}
\begin{array}{rcl}
< r_{13}^2+r_{12}^2 >&=&\sum_{K K^{\prime} l_{x_{1}} L S} {
{\displaystyle{\int}}_0^{\infty}} \rho 
^2 d\rho U_{K l_{x_{1}}LS}(\rho) U_{K^{\prime} l_{x_{1}}LS}(\rho) {
{\displaystyle{\int}}_0^{\pi /2}} 
{^{(2)}P_{K}}^{l_{x_{1}}, l_{x_{1}}}(\Phi) {^{(2)}P_{K^{\prime}}}^{l_{x_{1
}}, l_{x_{1}}}(\Phi)\\ 
&\times& \left[\frac{1}{2 a_{23}^{2}}~ cos^2\Phi + \frac{2}{a_{(23)1}^{2}}
~ sin^2\Phi\right] cos^2\Phi
sin^2\Phi d\Phi.
\end{array}
\end{equation}
The rms separation between the valence $\Lambda$-hyperons
$(R_{\Lambda\Lambda})$ is given by the expression 
\begin{equation}
R_{\Lambda\Lambda}=\left[ < r_{23}^2 > \right]^{1/2},
\end{equation}
where 
\begin{equation}
\begin{array}{rcl}
< r_{23}^2 >&=&\frac{1}{a_{23}^{2}} \sum_{K K^{\prime} l_{x_{1}} L S}
{ {\displaystyle{\int}}_0^{\infty}} \rho 
^2 d\rho U_{K l_{x_{1}} L S}(\rho) U_{K^{\prime} l_{x_{1}} L S}(\rho)\\
&\times& { {\displaystyle{\int}}_0^{\pi /2}} {^{(2)}P_{K}}^{l_{x_{1}},
l_{x_{1}}}(\Phi) 
~{^{(2)}P_{K^{\prime}}}^{l_{x_{1}}, l_{x_{1}}}(\Phi) ~cos^4\Phi ~sin^2\Phi
d\Phi.
\end{array}
\end{equation}
The result of these calculations are recorded in Table IV. 
The variation of the ground state rms matter radius ($R_M$) with 
mass number (A) of the double-$\Lambda$ hyper nuclei 
is displayed in Fig. 10. 

\begin{figure}
\centering
\fbox{\includegraphics[width=0.75\linewidth, height=0.6\linewidth]{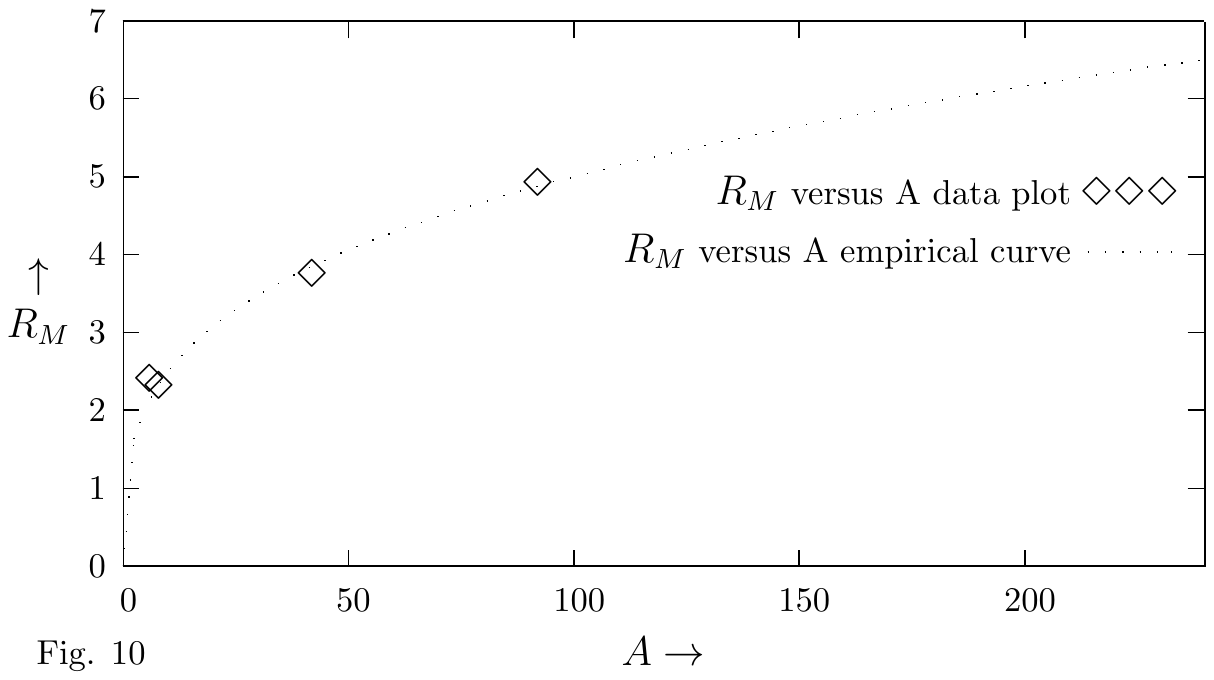}}
\caption{Variation of ground state rms matter radius $R_M$ (fm) with mass A (in nucleon mass unit) of the double-$\Lambda$ hypernuclei}
\label{fig:boxed_graphic}
\end{figure}
Finally we have calculated the contribution of various orbital angular 
momenta $l_{x_{1}}$ at $K_{max}=20$ to the ground state probability 
density distribution to identify the dominating component of 
the partial waves, and the results are presented in Table V.

\section {\bf Summary and Conclusion}
Since hyperons and nucleons both have three quark (qqq) structure (eg.
$p\rightarrow uud$, $n\rightarrow udd$, $\Lambda^{0}\rightarrow uds$ etc.)
interactions among hyperons as well as with nucleons should give
valuable inputs to the knowledge of strong interactions. But not much
attention has so far been directed to the study of hyperon-hyperon and
hyperon-nucleon interaction through the critical investigation of 
hypernuclei. We have undertaken a systematic study of the bound state 
properties of few light and medium mass hypernuclei to gather some 
idea about nature of hyperon-hyperon and hyperon-nucleon interactions 
and the internal structure of the three-body systems. The hyperspherical 
harmonics expansion (HHE) method 
adopted here is an essentially exact method, where calculations can be 
carried out up to any desired precision by gradually increasing the 
expansion basis. For $_{\Lambda\Lambda}^{6}$He, the calculated 
two-$\Lambda$ separation energy $B_{\Lambda\Lambda}$ 
and $\Lambda\Lambda$ bond energy $\Delta B_{\Lambda\Lambda}$ 
(see Table III) both at $K_{max}=20$ are well within the 
experimental error limits $7.25 \pm 0.19 MeV$ [7] and $1.01 MeV$ [7].
Both values are less than our previous calculation [52] done by using 
old data of [5]. 
The strength of the core-$\Lambda$ potential for a fixed range 
parameter ($\eta$), is found to become weaker as the mass of the core 
increases. And this trend is found to fit a logarithmic function 
of the form \begin{equation} v_0(A_c)=\delta_1 \log_e(A_c)+
\delta_2\end{equation} with $\delta_1=1.55033 MeV$ and 
$\delta_2=-43.8843 MeV$. In the low mass region $v_0$ decreases 
faster than in the mid mass range before achieving saturation in the
heavier mass region. This phenomenon may be viewed as the effect of 
screening of the interacting nucleon embedded in the core, by 
the surrounding nucleons. And this effect of screening increases 
as the number of neighboring nucleons increases. One may expect 
such a result from physical intuition. 
Following the above empirical formula one can estimate the depth of
the effective $\Lambda N$ potential for the single-$\Lambda$ hypernuclei
in any mass range. The single $\Lambda$ separation energy $B_{\Lambda}$
is found to increase with the increase in core mass approximately
following the equation \begin{equation} B_{\Lambda}(A_c)=\delta_3 
log_e(A_c)+\delta_3 \end{equation} with $\delta_3=6.21483 MeV$ and 
$\delta_4=-5.55812 MeV$. Similarly, the trend of variation of the
two-$\Lambda$ separation is found to follow the equation 
\begin{equation} B_{\Lambda\Lambda}(A)=\delta_5 log_e(A) 
+\delta_6\end{equation} with $\delta_7=12.3892 MeV$ and 
$\delta_8=-11.2555 MeV$. Thus Eq. (42) and (43) may respectively be useful
in determining $B_{\Lambda}$ and $B_{\Lambda\Lambda}$ for any one- and two
$\Lambda$ hypernuclei. Also the approximate size of any double $\Lambda$ 
hypernuclei of mass number (A) can be estimated by our empirical formula 
\begin{equation}R_M(A)=\delta_7 A^{\delta_8}\end{equation}
 with $\delta_7=1.25244 fm$ and $\delta_8=0.300609 fm$. 
It can be seen from energy level diagram (see Fig. 9) that the energy 
levels for the exotic isobars $\Lambda\Lambda^8$He and 
$\Lambda\Lambda^8$Be are almost identical. However the slight difference
in their one- and two-$\Lambda$ separation energies (see Column 4 of 
Table II and III) is due to the relatively stronger Coulomb repulsion
in the core of the latter. The calculated values of the partial
probability (see Table V) shows that  
larger contribution to the ground state total wave
function comes from the $l_{x_{1}}=0$ partial wave and 
the rest cones from the higher even $l_{x_{1}}$ components. The 
contribution from the odd $l_{x_{1}}$ values are zero.  From the calculated values of the single-
and double- $\Lambda$ separation energies (see Table II \& III) and the
rms matter radii (see Table IV) it may be concluded that the addition 
of one- or two-$\Lambda$ particle to the core makes the system strongly
bound and more compact. This is quite true as the weakly bound $^6He$ 
or unbound $^6Be$ core becomes strongly bound on addition of one or two 
$\Lambda$ hyeperon(s). Hence, it is quite possible that our predictions 
for the binding energies of the ground and excited states of
experimentally unobserved double $\Lambda$ hypernuclei are valid. Thus we
may say that that the $\Lambda N$ force is much stronger and much
attractive than the $NN$ force. Perhaps this may be due to the presence of
the strange quark in the $\Lambda$ hyperon quark constituents.\\ 

{\bf Acknowledgments}
The author gratefully acknowledges computer facilities provided by the Department of Physics, Aliah University.

\section{Figure Caption}
\begin{enumerate}
\item {\bf  Fig. 1} Choice of Jacobi coordinates for the partition 1.

\item {\bf Fig. 2}  Plot of: (i) $\Lambda\Lambda$ potential; 
(iia) effective $\Lambda N$ potential for $_{\Lambda\Lambda}^{6}$He;
(iib) effective $\Lambda N$ potential for 
$_{\Lambda\Lambda}^{92}$Zr; (iiia) core-$\Lambda$ folded potential for
$_{\Lambda\Lambda}^{6}$He; 
(iiib) core-$\Lambda$ folded potential for $_{\Lambda\Lambda}^{92}$Zr; 
(v is in MeV and r is the relative separation)

\item{\bf  Fig. 3}  Plot of lowest eigen potential $\omega_0(MeV)$ against hyperradial distance $\rho$ for double-$\Lambda$ hypernuclei 
(a) $_{\Lambda\Lambda}^{6}$He, (b) $_{\Lambda\Lambda}^{42}$Ca and  (c) $_{\Lambda\Lambda}^{92}$Zr

\item {\bf Fig. 4}  Plot of core-$\Lambda$ potential depth $v_{c\Lambda}$(MeV) against mass of core $(A_c)$

\item {\bf Fig. 5} Plot of ground  and excited state wavefunctions for $_{\Lambda\Lambda}^{6}$He

\item {\bf Fig. 6} Plot of ground  and excited state wavefunctions for $_{\Lambda\Lambda}^{42}$Ca

\item{\bf  Fig. 7}  Plot of ground  and excited state wavefunctions for $_{\Lambda\Lambda}^{92}$Zr

\item {\bf Fig. 8} Plot of ground state- (i) Single-$\Lambda$ separation energy $B_{\Lambda}$(MeV) and (ii) Two-$\Lambda$ separation 
energy $B_{\Lambda\Lambda}$ (MeV), against mass of the core $A_c$

\item {\bf Fig. 9} Energy Level diagram of  Double-$\Lambda$ Hypernuclei [$B_{\Lambda\Lambda}$ is in MeV]

\item{\bf  Fig. 10} Variation of ground state rms matter radius $R_M$ (fm) with mass A (in nucleon mass unit) of the doule-$\Lambda$ hypernuclei
\end{enumerate}

\section{Tables}
\begin{table}
\begin{center}
{\bf Table I. Parameters of  ${\Lambda\Lambda}$ interaction.}\\
\vspace{5pt}
\begin{tabular}{|c|c|c|c|}  \hline
$i \rightarrow$ &1 &2&3 \\  \hline   
$\beta_{i} (fm)$ &1.5 &0.90&0.5 \\  \hline   
$V_{i} (MeV)$&-8.967&-142.385&880.700 \\ \hline
\end{tabular}
\end{center}
\end{table}
\begin{table}
\begin{center}
{\bf Table II. Parameters of the $\Lambda N$ potential and corresponding
$\Lambda$ separation energy in the core-$\Lambda$ subsystems.}\\
\vspace{5pt}
\begin{tabular}{|c|c|c|c|c|}\hline
System&\multicolumn{2}{|c|}{$\Lambda$-N potential} 
&\multicolumn{2}{|c|}{$B_{\Lambda} (MeV)$} \\
\cline{4-5}
&\multicolumn{2}{|c|}{parameters}&\multicolumn{2}{|c|}{Ground state}
\\
\cline{2-5}
&$V_{0} (MeV)$&$\chi (fm)$& Expt.&Calc.\\\hline 
$_{\Lambda}^{5}$He&-43.6102&1.41&$3.12\pm 0.02 ~ [12,23]$&3.1200 \\\hline  
$_{\Lambda}^{7}$He&-39.6328&1.41&$5.23\pm 0.00 ~ [22]$&5.2300 \\ \hline  
$_{\Lambda}^{7}$Be&-39.3650&1.41&$5.16\pm 0.08 ~ [23]$&5.1600 \\ \hline  

$_{\Lambda}^{33}$S &-38.0436&1.41&$17.96\pm 0.00 ~ [24]$&17.9600 \\ \hline  
$_{\Lambda}^{41}$Ca&-37.8497&1.41&$19.24\pm 0.00 ~ [24]$&19.2400 \\ \hline  
$_{\Lambda}^{91}$Zr&-36.3328&1.41&$22.10\pm 0.30 ~ [25]$&22.1000 \\ \hline  
\end{tabular}
\end{center}
\end{table}
\begin{table}
\begin{center}
{\bf Table III. Two-$\Lambda$ separation energy in the ground and
excited\\ states of different double-$\Lambda$ hypernuclei at
$K_{max}=20$.}\\ 
\vspace{5pt}
\begin{tabular}{|c|c|c|c|c|c|}  \hline
Hypernuclei&Bound State &\multicolumn{2}{|c|}
{$B_{\Lambda\Lambda} (MeV)$}&\multicolumn{2}{|c|}{$\Delta
B_{\Lambda\Lambda} (MeV)$}\\  
\cline{3-6}&$J^{\pi}$&Expt.&Calc.&Expt.&Calc. \\\hline
$_{\Lambda\Lambda}^{ 6}$He&$0^+$&7.25$\pm$0.19[4]&7.2501& 1.01[4]
& 1.0101\\ 
&$0^+_1$&       -      &0.8613&-&  \\\hline 
$_{\Lambda\Lambda}^{8}$He&$0^+$&             -&11.5815&  -  & 1.1215\\ 
&$0^+_1$&       -      &2.7056&-&  \\\hline
$_{\Lambda\Lambda}^{8}$Be&$0^+$&             -&11.4344&  -  & 1.1144\\ 
&$0^+_1$&       -      &2.6451&-&  \\\hline
$_{\Lambda\Lambda}^{34}$S&$0^+$&             -&36.7957&  -  & 0.8757\\ 
&$0^+_1$&       -      &24.4038&-& \\
&$0^+_2$&       -      &14.1298&-& \\
&$0^+_3$&       -      &07.9952&-& \\
&$0^+_4$&       -      &02.1448&-& \\\hline
$_{\Lambda\Lambda}^{42}$Ca&$0^+$&             -&39.2357&  -  & 0.7557\\ 
&$0^+_1$&       -      &27.5745&-& \\
&$0^+_2$&       -      &17.0814&-& \\
&$0^+_3$&       -      &10.2213&-& \\
&$0^+_4$&       -      &04.3325&-& \\\hline
$_{\Lambda\Lambda}^{92}$Zr&$0^+$&             -&44.5463&  -  & 0.3463\\ 
&$0^+_1$&       -      &36.1553&-& \\
&$0^+_2$&       -      &27.0817&-& \\
&$0^+_3$&       -      &18.6847&-& \\
&$0^+_4$&       -      &11.4110&-& \\
&$0^+_5$&       -      &03.0006&-& \\\hline
\end{tabular}
\end{center}
\end{table}
\begin{table}
\begin{center}
{\bf Table IV. The r.m.s. matter radii of the ground and excited
states of different double-$\Lambda$ hypernuclei at $K_{max}=20$.}\\ 
\vspace{5pt}
\begin{tabular}{|c|c|c|c|c|c|}  \hline
Hypernuclei&Bound
State&$R_{A}(fm)$&$R_{c\Lambda}(fm)$&$R_{\Lambda\Lambda}(fm)$ 
&$R_{(\Lambda\Lambda)c}(fm)$\\
&$J^{\pi}$&&&&\\\hline
$_{\Lambda\Lambda}^{ 6}$He&$0^+$&2.3842&3.1804&4.1437&2.4130\\ 
&$0^+_1$&2.7823&3.9536&5.2301&2.9652 \\ \hline 
$_{\Lambda\Lambda}^{ 8}$He&$0^+$&2.2939&2.9081&3.8697&2.1711\\
&$0^+_1$&2.4952&3.4427&4.6361&2.5453 \\\hline 
$_{\Lambda\Lambda}^{ 8}$Be&$0^+$&2.2970&2.9167&3.8810&2.1775\\
&$0^+_1$&2.5005&3.4560&4.6542&2.5552 \\\hline
$_{\Lambda\Lambda}^{34}$S &$0^+$&3.4592&2.9783&4.0864&2.1670\\
&$0^+_1$&3.4666&3.0998&4.2331&2.2647 \\
&$0^+_2$&3.4840&3.3706&4.6237&2.4528 \\
&$0^+_3$&3.5095&3.7331&5.1750&2.6909 \\
&$0^+_4$&3.5377&4.0998&5.7564&2.9196 \\\hline
$_{\Lambda\Lambda}^{42}$Ca&$0^+$&3.7286&3.1140&4.2798&2.2623\\
&$0^+_1$&3.7332&3.2124&4.3931&2.3441 \\
&$0^+_2$&3.7450&3.4484&4.7293&2.5099 \\
&$0^+_3$&3.7628&3.7779&5.2220&2.7305 \\
&$0^+_4$&3.7830&4.1209&5.7556&2.9496 \\\hline
$_{\Lambda\Lambda}^{92}$Zr&$0^+$&4.9045&3.8392&5.3125&2.7720\\
&$0^+_1$&4.9049&3.8581&5.3148&2.7969 \\
&$0^+_2$&4.9078&4.0007&5.5133&2.8994 \\
&$0^+_3$&4.9124&4.2110&5.8141&3.0466 \\
&$0^+_4$&4.9179&4.4546&6.1672&3.2147 \\
&$0^+_5$&4.9230&4.6693&6.4821&3.3612 \\\hline
\end{tabular}
\end{center}
\end{table}

\begin{table}
\begin{center}
{\bf Table V. The contribution of the orbital angular momenta $l_{x_{1}}$ to the probability density in the ground state of  
  double $\Lambda$-hypernuclei at  $K_{max}=20$.}\\
\vspace{5pt}
\begin{tabular}{|c|c|c|c|c|c|c|c|}  \hline
Hypernuclei&\multicolumn{5}{|c|}{Partial probability $P_{l_{x_{1}}}$ for
$l_{x_{1}}=$}\\ 
\cline{2-6}
&0&1&2&3&4 \\ \hline
$_{\Lambda\Lambda}^{ 6}$He&0.996130&0.000000&0.003854&0.000000&0.000016\\
\hline 
$_{\Lambda\Lambda}^{8}$He&0.997786&0.000000&0.002210&0.000000&0.000003\\
\hline 
$_{\Lambda\Lambda}^{8}$Be&0.997757&0.000000&0.002240&0.000000&0.000004\\
\hline 
$_{\Lambda\Lambda}^{34}$S&0.999900&0.000000&0.000099&0.000000&0.000001\\
\hline 
$_{\Lambda\Lambda}^{42}$Ca&0.999878&0.000000&0.000121&0.000000&0.000001\\
\hline 
$_{\Lambda\Lambda}^{92}$Zr&0.999744&0.000000&0.000254&0.000000&0.000002\\
\hline 
\end{tabular}
\end{center}
\end{table}


\newpage 
\begin{thebibliography}{References:}
\bibitem{d}  M. Danysz et al., Nucl. Phys. {\bf 49}, 121(1963).
\bibitem{p} D. J. Prowse, Phys. Rev. Lett. {\bf 17}, 782(1966).
\bibitem{ddfmpz} R. H. Dalitz, D. H. Davis, P. H. Fowler, A. Montwill, J.
Poriewski, and J. A. Zakrzewski, Proc. R. Soc. London, Ser. {\bf A426},
1(1989).
\bibitem{dd}D. H. Davis, Nucl. Phys. {\bf A754} 3c(2005); 
R. H. Dalitz, Nucl. Phys. {\bf A754} 4c(2005).
\bibitem{him} H. Himeno et al., Prog. Theo. Phys. 
{\bf Vol. 89}, No. 1, 109(1993).
\bibitem{m} L. Majling, Nucl. Phys. {\bf A585}, 211c(1995).
\bibitem{t} Takahashi et. al., Phys. Rev. Lett. {\bf 87}, 212502(2001).
\bibitem{a} S. Aoki et al., Prog. Theo. Phys. {\bf 85}, 1287(1991).
\bibitem{nsfn} H. Nemura, Y. Suzuki, Y. Fuziwara and C. Nakamoto, Prog.
Theo. Phys.  
{\bf Vol. 105}, No. 5, 929(2000).
\bibitem{rnc} R. R. Roy and B. P. Nigam, Nuclear Physics Theory and
Experiment, Fifth Wiley Eastern Reprint, February (1993)pp-389; J. Caro et
al., Nucl. Phys. {\bf A646}, 299(1999). 
\bibitem{Yama} Y. Yamamoto et al., Nucl. Phys. {\bf A547}, 233c(1992).
\bibitem{g} A. Gal, Advances in Nucl. Phys., eds. M. Baranger and E. Vogt.
{\bf Vol.8}, 1(1975, Plenum).
\bibitem{a} G. Alexander et al., Phys. Rev. {\bf 173}, 1452(1968).
\bibitem{sz} B. Sechi - Zorn et al., Phys. Rev. {\bf 175}, 1735(1968).
\bibitem{k} J. A. Kadyk et al., Nucl. Phys. {\bf B27}, 13(1971).
\bibitem{h} J. M. Hauptman, LBL Report No. {\bf LBL-3608}(1974).
\bibitem{e} F. Eisele et al., Phys. Lett. {\bf 37B}, 204(1971).
\bibitem{e} R. Engelmann et al., Phys. Lett. {\bf 21}, 487(1966).
\bibitem{th} Y. C. Tang and R. C. Herndon, Nuovo Cimento
{\bf XLVIB}, N.1., 117(1966).
\bibitem{dr} R. H. Dalitz and G. Rajasekaran, Nucl. Phys. {\bf 50},
450(1964). 
\bibitem{ab} S. Ali and A. R. Bodmer, Phys. Lett. {\bf 24B}, 343(1967).  
\bibitem{buc} A. R. Bodmer, Q. N. Usmani, and J. Carlson, Nucl. Phys. {\bf
A422}, 510(1984).
\bibitem{bmz} H. Bando, T. Motoba and Z. J. Zofka, Int. J. Mod. Phys. 
{\bf A5}, 4021(1990); see references therein.
\bibitem{lgm} G. A. Lalazissis, M. E. Grypeos, and S. E. Massew, Phys. 
Rev. {\bf C37}, 2098(1988).
\bibitem{ymhin} Y. Yamamoto, T. Motoba, H. Himeno, K. Ikeda, and S.
Nagata, Prog. Theor. Phys. Suppl. {\bf 117}, 361(1994).
\bibitem{cdl} C. D. Lin, Phys. Rep. {\bf 257}, 1(1995).
\bibitem{g}T. H. Gronwall, Phys. Rev. {\bf 51}, 655(1937).
\bibitem{m}J. Macek, J. Phys. {\bf B 1}, 831(1968).
\bibitem{bn}J. L. Ballot and J. Navarro, J. Phys.{\bf B 8}, 172(1975).
\bibitem{efm}V. D. Efros, A. M. Frolov and M. I. Mukhtarova, J. Phys.{\bf
B 15}, 1819(1982). 
\bibitem{dcm} T. K. Das, R. Chattopadhyay and P. K. Mukherjee, Phys. Rev.
\underline{A50}, 3521(1994).
\bibitem{cd} R. Chattopadhyay and T. K. Das, Phys. Rev. {\bf A56},
1281(1997). 
\bibitem{kdd} Md. A. Khan, S. K. Datta, and T. K. Das, FIZIKA B
(Zagreb){\bf 8}, 4, 469(1999).
\bibitem{dcf}  T. K. Das, H. T. Coelho and M. Fabre de la Ripelle,  Phys.
Rev. {\bf C26}, 2288(1982).
\bibitem{kddp} Khan M. A., Dutta S. K., Das T. K. and Pal M. K., J. Phys.
G: Nucl. Part. Phys. {\bf 24}, 1519(1998). 
\bibitem{s} Yu. A. Simonov, Yad. Fiz. {\bf 3}, 630(1960) [Sov. J.
Nucl. Phys. {\bf 3}, 461(1960)]; in proced. of the international
symposium on the present status and novel developments in the nuclear many
body problem. Rome, 1972, edited by F. Calogena and C. Ciofi Degli Atti
(Editrice composition, Bologna, 1973), p527; Sov. J. Nucl. Phys.
{\bf 7}, 722(1968).
\bibitem{zb} F. Zernike, H. C. Brinkman, Proc. Kon. Acad. Wtensch,
{\bf 33}, 3(1975).
\bibitem{f}M. Fabre  de la Ripelle, Proc. Int. Sch. Nucl. Theo. Phys.,
Predeal 1969.
\bibitem{f} M. Fabre de la Ripelle, Comp. Reend. Acad. Sci.,
{\bf 269B}, 80(1970); {\bf 273A}, 1007(1971).
\bibitem{bf} J. L. Ballot and M. Fabre de la Ripelle, Ann. Phys. ( N. Y.)
  {\bf 127}, 62(1980).
\bibitem{r} J. M. Richard, Phys. Rep. {\bf 212}, 1(1992).
\bibitem{lfgsl} H. Leeb, H. Fiedeldey, E. G. O. Gavin, S. A. Sofianos
 and R. Lipperheide, Few Body Systems {\bf 12}, 55(1992).
\bibitem{bn} N. Barnea and A. Novoselsky, Ann. Phys. ( N. Y.)
  {\bf 256}, 192(1997).
\bibitem{whk} S. Watanabe, Y. Hosoda and D. Kato, J. Phys. {\bf B26},
L495(1993). 
\bibitem{cks} F. Cooper, A. Khare and U. Sukhatme, Phys. Rep. {\bf 251},
267(1995). 
\bibitem{dcf} T. K. Das, H. T. Coelho and M. Fabre de la Ripelle, Phys.
Rev. {\bf C26}, 2281(1982).
\bibitem{dl}R. H. Dalitz and L. Liu, Phys. Rev {\bf 116}, 1312(1959).
\bibitem{bd} M. M. Block and R. H. Dalitz, Phys. Rev. Lett. {\bf 11},
96(1963). 
\bibitem{a} J. B. Adams, Phys. Rev. {\bf 156}, 1611(1967).
\bibitem{chk} C. Y. Chung, D. P. Heddle and L. S. Kisslinger, Phys. Rev.
{\bf C27}, 335(1983). 
\bibitem{kg} B. H. Mc Kellar and B. F. Gibson Proc. Int. Conf. on
Hypernuclear and Kaon Physics, 1982, Heidelberg, ed. B. Povh, p-156;
Phys. Rev. {\bf C30}, 322(1984).
\bibitem{kd} M. A. Khan and T. K. Das, Pramana- J. Phys, {\bf 56} No.1,
57(2001).
\end{thebibliography}
\end{document}